 \RequirePackage{lineno} 
 \RequirePackage{fix-cm}
\RequirePackage{amsmath}
\documentclass[12pt]{article}
\usepackage{amssymb}
\usepackage{amssymb,amsfonts}

\usepackage{amsmath,amssymb}
\usepackage{tikz}
\usetikzlibrary{decorations.pathreplacing}

\usepackage{ytableau}
\usepackage{tikz}
\usepackage{tikz-cd}

\usepackage{pdflscape}
\usepackage{graphicx}
\usepackage{mathtools}
\usepackage{amsthm}
\usepackage{setspace}
\usepackage{diagbox}
\usepackage{adjustbox}
\usepackage{array}
\usepackage{multirow}
\usepackage{booktabs}
\usepackage[all,arc]{xy}
\usepackage{enumerate}
\usepackage{mathrsfs}
\usepackage{setspace}
\newtheorem{theorem}{Theorem}
\newtheorem{definition}{Definition}

\newtheorem{proposition}{Proposition}

\theoremstyle{definition}

\theoremstyle{remark}

\makeatletter
\let\c@equation\c@thm
\makeatother
\numberwithin{equation}{section}

\title{Formation of coalition structures as a  non-cooperative game}

\author{Dmitry Levando, \thanks{ 
Moscow, Russian Federation, independent. No funds, grants, or other support was received.  
Acknowledge:  Nick Baigent, Phillip Bich,   Giulio Codognato, Ludovic Julien,  Izhak Gilboa, Olga Gorelkina,  Mark Kelbert, Anton Komarov,  Roger Myerson, Ariel Rubinstein,  Shmuel Zamir. Many thanks for  advices and for discussion to participants of  SIGE-2015, 2016 (an earlier version of the title was ``A generalized Nash equilibrium"), CORE 50  Conference,  Workshop of the Central European Program in Economic Theory, Udine (2016), Games 2016 Congress, Games and Applications, (2016) Lisbon, Games and Optimization at St Etienne (2016). All possible mistakes are only mine. 
 \newline E-mail for correspondence: dlevando (at)  hotmail.ru.
}}
\date{} 

\begin{document}

\maketitle
\begin{abstract}

 We study coalition structure formation with intra and inter-coalition externalities in the introduced  family of nested non-cooperative simultaneous finite games.
A non-cooperative game  embeds a coalition structure formation mechanism, and  has two outcomes: an allocation of players over coalitions and a payoff for every player. Coalition   structures  of a game are  described by Young diagrams.  They serve to enumerate coalition structures and  allocations of players over them. For every coalition structure  a player has a  set of finite strategies. A player chooses a coalition structure and a strategy.  

A (social) mechanism eliminates conflicts in individual choices and produces final coalition structures. Every final coalition structure is a non-cooperative game. Mixed equilibrium  always exists and consists of  a  mixed strategy profile, payoffs and  equilibrium coalition structures.  We use a maximum coalition size  to parametrize  the family of the games. 
The non-cooperative game of Nash   is a partial case of the  model.   The result is different from  the Shapley value, a strong Nash, coalition-proof equilibria,  core solutions, and   other equilibrium concepts.   We supply few  non-cooperative  coalition structure stability criteria. 
 \end{abstract}


\noindent{\bf Keywords: } Noncooperative Games, cooperative games, Nash Program, group formation, mechanism design.

\noindent{\bf JEL} :   C71, C72, D02, D71 

\section{Introduction}

The  paper was inspired by John Nash{'}s   \textquotedblleft Equilibrium Points in n-person games, \textquotedblright  (1950). This remarkably short, but highly influential note of only  5 paragraphs  established an equilibrium concept and the proof of its existence  without  explicit specification of a final coalition structure. Prior to  the Nash{'}s paper, the generalization of the concept of equilibrium for coalition  games provided by von Neumann for the case of two-players zero-sum game was done by portioning the players into two groups and regarding several players as a single player.   However up to now, none of these  competing approaches   revealed expected  progress in  non-cooperative   multiple group formation.     ept for this game?

Since those times the game theory landscape  is exposed  for    the notorious  dichotomy:  the cooperative game theory (CGT) versus  the non-cooperative game theory (NGT). CGT deals with coalitions as elementary items with vague individual activity, NGT deals with strategic individual behavior;  the theories are compared in Table \ref{differences}.   The research program of the current paper is in the third column:   it targets  goals of CGT by the tools of NGT. We study a non-cooperative construction of multiple coalitions from  a non-cooperative setup, with existence of    intra and inter coalition externalities, caused by individual actions of players.   The   paper offers  a  generalization of the non-cooperative game from Nash (1950) to address the problem  of  multiple coalition  formation absent  in Nash (1953). We present a simultaneous, finite strategies, non-cooperative game, which  has  (mixed)  equilibrium coalition structures. 
The difference  of the approach with CGT is, that  payoff are  not ``awarded to each player" (Serrano, 2020), but  emerge from actions of all  interacting players,  even from those in other coalitions.

\begin{table}[htp]
\caption{Comparison  of game theories and the suggested research program}
\begin{center}
\begin{tabular}{|c|c|c|c|}
\hline
 & CGT & NGT & this paper \\
 \hline
type of a  mapping & set to point &  vector  to   vector  & tensor to tensor  \\ \hline
 individual action & no &  always  &  always \\ \hline
 individual motivation &no & always  &  always  \\ \hline
 \begin{tabular}{c} explicit allocation \\ of players over coalitions \end{tabular} & yes &  no  & yes \\ \hline \begin{tabular}{c} 
 inter-coalition  \\ externalities \end{tabular}  &  sometimes &  vague  & \begin{tabular}{c} explicit, from \\  all players  in \\ other coalitions \end{tabular} \\ \hline
\begin{tabular}{c}   intra-coalition \\  externalities   \end{tabular}& no & \begin{tabular}{c} explicit, but   \\ within one  \\ implicit coalition \end{tabular}  &   \begin{tabular}{c} explicit, from  \\ players in  the \\ coalition  \end{tabular}  \\
 \hline
  \end{tabular}
\end{center}
\label{differences}
\end{table}%

 
 Importance  of studying  externalities between coalitions, a case impossible within  the Nash Program, was mentioned by Maskin, (2011).
Practice of social   analysis and social design also  requires studying non-cooperative  and simultaneous formation of \textit{multiple} coalitions, or coalition structures. This makes Nash program to construct one coalition from individual actions be too  restrictive.  

We will use the terms   ``coalition structure," or  a ``partition,"\footnote{Existing literature uses both terms.}  to denote  a collection of non-overlapping subsets from a set of players, which  in a union  make the original set. A group, or a coalition, is an element of a coalition structure or  of  a  partition. 

A partition induces two types of effects on a player's payoff. First, through actions of players of  the same coalition, what  producers  \textit{intra}-coalition externalities  inside the coalition. 
Second, from the players who  are outside  the  coalition, and  produce   \textit{inter}-group externalities for other coalitions. 

 The paper of Nash (1953)  suggested that cooperation   should be studied  using non-cooperative fundamentals with formation  of only  one group,   this conjecture  known nowadays  as the Nash Program.  His approach  includes  an  \textquotedblleft umpire\textquotedblright \footnote { Nash (1953): \textquotedblleft The point of this discussion is
that we must assume there is an adequate mechanism for forcing the
players to stick to their threats and demands once made, for one to
enforce the bargain once agreed. Thus, we need a sort of umpire who
will enforce contracts or commitments.\textquotedblright} that filters\footnote{i.e. checks that actions satisfy some rules, known as the axioms.}   actions of two  players to  let them reach an equilibrium.\footnote{The paper has also restriction on construction of payoffs.}  Serrano (2004) commented this as: \textquotedblleft the vexed issue of
enforcement of outcomes cannot be overlooked, and one must assume
either an enforcement by the designer or by other vehicles which will typically be left unspecified.\textquotedblright  
  Our approach    exploits  an outside mechanism,   which serves to describe ``social artifacts'' of coalition formation.  
 
The best analogy for a difference between   the Nash Program and   the current research is the  difference between partial and general equilibrium analysis in economics. The former isolates one market  ignoring cross-market interactions, the latter explicitly studies cross market interactions, along with activities  at every specific market.

Formation and disintegration  of  every coalition   depends on individual expectation of it{'}s members from every feasible alternative to deviate, including iterative reasoning about expectations of all   players, including those in other coalitions.  In the suggested model  a player has alternatives, described in terms of formed coalition structures and allocations of other players over them. And the Young diagrams serve to enumerate the appearing alternatives.

The mechanism of our model is the following.  First, we impose a restriction  on  a maximum coalition size, and  using Young diagrams construct all possible coalition partitions   and all possible allocation of players. Second, for every possible allocation of players in  every possible Young diagram we   assign  every  player a  finite strategy set and payoff. Every coalition structure becomes  a playable  non-cooperative game. Third,    we  let players choose a coalition partition  and a strategy, from individual strategy set for  this coalition structure. Fourth, an external social mechanism\footnote{ we mean  the sociological term ``social artifacts.''} transforms  all individual choices into   final coalition structures and   individual strategies (they can differ from initial ones). The mechanism has an enforcement role to clear up conflicts of choices between players.  Every resulting coalition structure is a playable non-cooperative game with allocation of players over coalitions.  By topological reasons  an equilibrium   in mixed strategies always exists.   
 Increasing the size of a coalition we obtain a family of embedded games, where each has  coalition structures and an equilibrium, possibly in mixed strategies.

 The  model separates, those who make decisions,  agents, and social artifacts, which exist beyond agents. The latter include social norms,  customs and traditions,  brute force and enforcements along with a variety of   different transaction costs and imperfect information.  Agents adjust  individual interactive beliefs to those of other players, and operations of the mechanism. The mechanism is the link to the social design that  we do not study here. 
 
 The contributions of this paper are: a construction of a non-cooperative  simultaneous  game with a  coalition structure formation mechanism, and stability criteria, including   one  non-cooperative with self-enforcement for  the strong Nash equilibrium (Aumann, 1990).

 These results are different from the strong  Nash equilibrium, coalition-proof equilibrium, multiple value approaches of the cooperative game theory.    Existence of an equilibrium  does not depend on a number of deviators or their threats to deviate, but on individual actions of players. Non-cooperative  game of Nash is a partial case of the suggested model. 
 
 The paper has the following plan. Section 2 presents  an example approach, Section 3 presents  the model with another example, Section 4  introduces stability criteria, then follow  the Discussion and a Conclusion.

\section{An example of a game of two players}
The example generalizes   the Prisoner{'}s  Dilemma  for the case of explicit coalitions structures formation. It serves   to demonstrate, how the model  from  the next section  works.  The  example  argues  the popular statement,  that Pareto  efficient outcome of the Dilemma  means  cooperation as a membership in one coalition.  Cooperation in the example  is    creation of positive externalities, but not necessarily a belonging to the same coalition or  a group. To construct  coalition partitions  we demonstrate a simple mechanism, where  the grand coalition can be formed  only from  unanimous choices.   

\subsection{Matrix game: generalization of the Prisoners{'} dilemma}

There are 2 players, and they can form two types of coalition   structures. Letter $K$  is  a size of a maximum coalition  for any partition.   If  $K=1$   then there  is only one final partition,  $$\mathcal{P}(K=1) \equiv P_{separ}=\Big\{ \{ 1\},\{ 2\}\Big\},$$ with one player in each coalition.   For $K=2$ there are two final partitions,  $$ \mathcal{P}(K=2)=  (P_{separ},P_{joint} ) = \Big\{ \{ \{ 1\},\{ 2\}\} , \{ 1,  2 \}  \Big\},$$   where $P_{joint}=\{ 1, 2\}$ is the grand coalition. The partition structures $\mathcal{P}(K=1)$ and $\mathcal{P}(K=2)$ are nested, $\mathcal{P}(K=1) \subset \mathcal{P}(K=2)$.

We describe  sizes of coalitions and allocations of  players   in coalitions with the Young diagrams.  This allows us to partition a set of agents   by  sizes of coalitions.\footnote{ This section demonstrates the first   property of the Young diagrams, used in the model. The second one is presented  later.}   Everywhere in this model a    box contains  exactly  one player.  
 A coalition is a horizontal sequence of boxes, what in terminology  of the Young diagrams is a ribbon. Length of a ribbon is a size of the coalition. For $K=1$  there are  two ribbons length one, located vertically: \begin{ytableau}
    \none[1]  &   \\
    \none[1]  &   \\
  \none & \none[] 
\end{ytableau}.

For the maximum coalition size   $K=2$, there are  two cases. The case   $K=1$ is still valid here, but another case is feasible,  the grand coalition:   a horizontal ribbon length two, see   below, Figure \ref{separ_joint}.  
\begin{figure}
 \caption{An increase in available coalition structures with an increase in $K$.}
 \centering
 \label{separ_joint}
$P_{separ}$ \begin{tikzcd}
& {\begin{tabular}{c} $K=1$ \\ 
\begin{ytableau}
\none[1]  & \none[]   \\
\none[1] & \none[] 
\end{ytableau} 
\end{tabular}
}  \arrow[dl] \arrow[dr] & \\
{\begin{tabular}{c} 
\begin{ytableau}
\none[1]  & \none[]   \\
\none[1] & \none[] 
\end{ytableau} 
\end{tabular}
}  &separated<-K=2->joined& 
{\begin{ytableau}
\none[2]  &   \\ 
 \end{ytableau}  }; 
\end{tikzcd}$P_{joint}$
\end{figure}
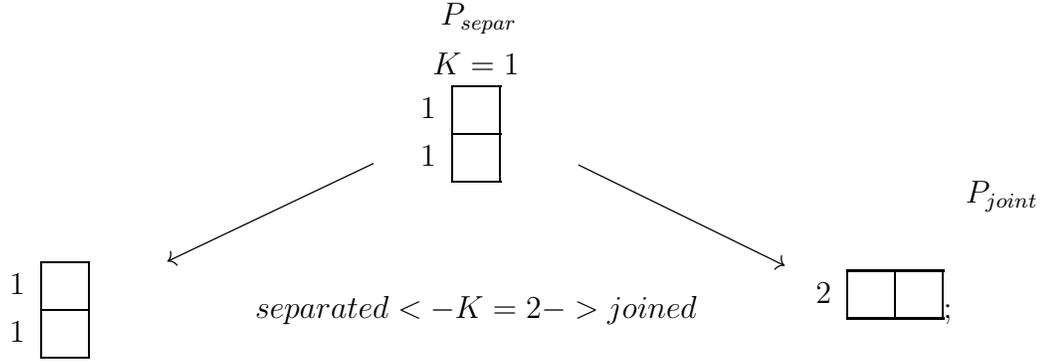

For  every   coalition structure a player has a set of strategies. For  $K=1$   player $i$, $i=1,2$, has two strategies: H(igh) and L(ow), $S_i(K=1)=(L_{i,K=1,P_{separ}}, H_{i,K=1,P_{separ}} )$.  
  Set of  strategies   of both players   for   $K=1$ is 
  $$S(K=1)= \times_{i=1,2}  \Big(L_{i,K=1,P_{separ}}, H_{i,K=1,P_{separ}} \Big) = \times_{i = 1,2} S_i(K=1).  $$
  
For   $K=2$   the coalition structure $P_{separ}$  and  the relevant strategies are   still feasible. Thus,  $L_{i,K=1,P_{separ}} = L_{i,K=2,P_{separ}}$ and $H_{i,K=1,P_{separ}} = H_{i,K=2,P_{separ}}$. But for $K=2$ there are additional strategies from  $P_{joint}$, so    $$S_i(K=2) =  \Big\{\underbrace{L_{i,K=2,P_{separ}}, H_{i,K=2,P_{separ}}}_{S_i(K=1) \mbox{ for } P_{separ}}, 
  \underbrace{L_{i,K=2,P_{joint}}, H_{i,K=2,P_{joint}}}_{\mbox{ + strategies for } P_{joint}}\Big\},$$  with a general element $s_i$.

The Young diagrams are implicit in the above construction of  $S_i(K=2)$, and we can write them explicitly:
$$S_i(K=2) =  \Big( S_i (P_{separ}) \Big) \cup \Big(S_i(P_{joint}) \Big),$$ where   $S_i (P_{k})  = \Big(L_{{i}, K=2, P_{k}},H_{{i}, K=2, P_{k}}\Big) $, $k \in \{P_{separ}, P_{joint} \}$. 
The  last notation for $S_i(K=2)$ will be used  for generalization of the model.  Then, a set of all  strategies of both players   for $K=2$ is
$$S(2) = S_1(K=2) \otimes  S_2(K=2) ,$$
with a general element $s$. We use the $\otimes$-product to combine all cases of individual strategy sets. 
Similar to partitions, strategy sets are nested  for different $K$, $S(K=1)  \subset S(K=2)$.

Tables \ref{default01}  collects all 16 outcomes of the game. Payoffs are constructed according to   the Prisoner{'}  Dilemma.
 One can observe   the nested structure of the  table  for  $K=1$ and $K=2$.

    {\small{\begin{table}[]
\caption{{Payoff  and coalition structures for the games: what players expect from a game. If players disagree, then outcomes are written separately. Pareto-efficient outcomes are underlined, equilibria are in the bold font.}   
} 
\begin{center}
\begin{tabular}{|c|c|c||c|c|} \hline
 & $L_{2,P_{separ}}$ & $H_{2,P_{separ}} $& $L_{2,P_{joint}} $& $H_{2,P_{joint}} $ \\  \hline
$L_{1,P_{separ}}$ & \underline{(0;0)}  $\colon \{\{1\},\{ 2\}\}$  &(-5;3)$\colon \{\{1\},\{ 2\} \}$ &
 \begin{tabular}{c}\underline{(0;0)} \\ $\colon\{\{1\},\{ 2\} \}$ \\ \hline  \underline{(0;0)}  \\ $\colon \{1, 2 \}$ \end{tabular}& 
 
  \begin{tabular}{c}(-5;3) \\  $\colon \{\{1\},\{ 2\} \}$  \\ \hline  (-5;3) \\ $\colon \{1,2\}$  \end{tabular}

\\ \hline
$H_{1,P_{separ}}$ & (3;-5) $\colon \{\{1\},\{ 2\}\}$ & \textbf{(-2;-2)$\colon \{\{1\},\{ 2\}\}$} & 
 \begin{tabular}{c}(3;-5) \\ $\colon \{\{1\},\{2\} \}$ \\ \hline  (3;-5) \\ $\colon \{1,2\}$ \end{tabular} &

 \begin{tabular}{c} \textbf{ (-2;-2)} \\  $\colon \{\{1\}, \{2\} \}$ \\ \hline   \textbf{ (-2;-2)} \\  $\colon \{1, 2 \}$ \end{tabular} \\ \hline \hline
 
$L_{1,P_{joint}}$ &  
\begin{tabular}{c} \underline{(0;0)}  \\ $\colon \{1,2\}$  \\ \hline  
\underline{(0;0)} \\ $\colon \{\{1\}, \{2 \}\}$
 \end{tabular} & 
  \begin{tabular}{c} (-5;3)  \\ $\colon\{1,2\}$  \\ \hline  
 (-5;3)  \\ $\colon \{\{1\},\{ 2\} \}$
 \end{tabular}
 & $  \underline{(0;0)} $ $\colon\{1,2\}$& $(-5;3)$  $\colon\{1,2\}$

  \\ \hline
$H_{1,P_{joint}}$ & 
 \begin{tabular}{c} (3;-5) \\ $\colon \{ 1,2\}$ \\ \hline  (3;-5) \\ $\colon \{\{1\},\{ 2\} \}$ 
 \end{tabular}
 & 
 \begin{tabular}{c}
\textbf{ (-2;-2)} \\ $\colon \{1,2\}$  \\  \hline \textbf{ (-2;-2)} \\ $\colon \{\{1\},\{ 2\} \}$ 
\end{tabular}
& (3;-5) $\colon \{1,2\}$ & \textbf{ (-2;-2) $\colon\{1,2\}$ } \\ \hline  
 \end{tabular}
\end{center}
\label{default01}
\end{table}%
 }}

 {\small{\begin{table}[]
\caption{{Payoff  and coalition structures for the games  with unanimous  formation  of the  grand coalitions}.  
{A cell contains a payoff profile and a coalition structure for the players \textit{after} an application of  the unanimous agreement mechanism $\mathcal{R}$. Pareto-efficient outcomes are underlined, equilibria are in the bold font.}
} 
\begin{center}
\begin{tabular}{|c|c|c||c|c|} \hline
 & $L_{2,P_{separ}}$ & $H_{2,P_{separ}} $& $L_{2,P_{joint}} $& $H_{2,P_{joint}} $ \\  \hline
$L_{1,P_{separ}}$ & \underline{(0;0)}  $\colon \{\{1\},\{ 2\} \}$  & (-5;3) $\colon \{\{1\},\{ 2\} \}$ & \underline{(0;0)}  $\colon \{\{1\},\{ 2\} \}$ &  (-5;3) $\colon \{\{1\},\{ 2\} \}$ \\ \hline
$H_{1,P_{separ}}$ & (3;-5)$\colon \{\{1\},\{ 2\} \}$& \textbf{(-2;-2)$\colon \{\{1\},\{ 2\} \}$} & (3;-5) $\colon \{\{1\},\{ 2\} \}$ &\textbf{ (-2;-2)  $\colon \{\{1\},\{ 2\} \}$} \\ \hline \hline
$L_{1,P_{joint}}$ & \underline{(0;0)} $\colon \{\{1\},\{ 2\} \}$& (-5;3) $\colon \{\{1\},\{ 2\} \}$&$  \underline{(0;0)} $ $\colon \{1, 2 \}$& $(-5;3)$  $\colon \{1,2\}$ \\ \hline
$H_{1,P_{joint}}$ & (3;-5)$\colon \{\{1\},\{ 2\} \}$& \textbf{ (-2;-2) $\colon \{\{1\},\{ 2\} \}$ }& $(3;-5) $ $\colon \{1,2\}$ & \textbf{ (-2;-2) $\colon \{1, 2\}$ }  \\ \hline  
 \end{tabular}
\end{center}
\label{default2}
\end{table}%
 }}

Now   we    compare  payoffs  and coalitions  for two strategy profiles:    $s=(L_{1,K=2,P_{separ}}, L_{2,K=2,P_{separ}})$  and  $s=(L_{1,K=2,P_{joint}}, L_{2,K=2,P_{joint}})$, which differ only in coalition structures.  Both  profiles lead to the  non-equilibrium Pareto-efficient outcomes. Traditional interpretation of    such outcome is    cooperation, with an assumption of a unanimous agreement to form a coalition. But  here, the strategy  profiles induce the same payoffs $(0,0)$  for \textit{different}  coalition structures:  
  $  \textbf{P}_{separ} = \Big\{  \{1\}, \{ 2\} \Big\}$ and 
  $\textbf{P}_{joint} =  \{ 1 ,  2\}.$
   Thus we can claim only that players generate posititive externalities for each other, but not necessarily stay in the same coalition.  
  
 \subsection{An example of a mechanism}
  There are cases, when players have conflicts in choosing  coalition structures. For example, one player chooses to be alone, $s_i=(L_{i,K=2,P_{separ}})$, and another chooses to be together $s_{-i}=(L_{-i,K=2,P_{joint}})$, and  then,  the players  disagree on  final coalition structures, see cells of Table \ref{default01}.\footnote{Note, that due to nested property  of strategy sets there is  $(L_{i,K=2,P_{separ}}) =(L_{i,K=1,P_{separ}})$.} So, non-cooperative coalition structure formation requires a resolution for the conflict.\footnote{a brief    discussion  of the approach  is  in Section 3.}

We describe a mechanism  to resolve the conflict as a mapping, and   for example, take  a  unanimous  mechanism  for coalition structures formation. 
 This means:  the players can be together, only if both choose to be together, in other words only if   both choose   $P_{joint} =\{1,2 \}$ and  are interested in the following Young diagram \begin{ytableau}
\none[2]  & &  \\ 
\end{ytableau}.  In all other cases the players  are allocated  to different coalitions  with the diagram \begin{ytableau}
\none[1]  &   \\ 
\none[1]  &   \\ 
\end{ytableau}.  

Given the mechanism, the players form the grand coalition,  only for the    strategy profiles  $ \{(Z_{i,K=2,\textbf{P}_{joint} }, \tilde{Z}_{-i, K=2,\textbf{P}_{joint}} )$, such that $Z, \tilde{Z} \in {H,L}$, $i = 1,2\}$. For all other strategy  profiles  only   partition $\textbf{P}_{separ} = \Big\{  \{ 1\}, \{ 2\}\Big\}$ is feasible for them, see  Figure \ref{defaultY} and Tables \ref{default2},  \ref{mappingR}.

We can describe a mechanism as a mappings of Young diagrams. see Figure \ref{defaultY}, where the first diagram is a choice of the first player, the second choice is one of the second, and at the right-hand part is a resulting  coalition structure.

\begin{figure}
\begin{center}
\caption{The unanimous rule in terms of the Young diagrams}
\label{defaultY}
\begin{tabular}{c}
$
\Big(\begin{ytableau}
    \none[1]  &   \\
    \none[1]  &   \\
  \none & \none[] 
\end{ytableau},  
\begin{ytableau}
    \none[1]  &   \\
    \none[1]  &   \\
  \none & \none[] 
\end{ytableau}
  \quad \Big)
\rightarrow   \begin{ytableau}
    \none[1]  &   \\
    \none[1]  &   \\
  \none & \none[] 
\end{ytableau}  $
\\
$
\Big(\begin{ytableau}
    \none[1]  &   \\
    \none[1]  &   \\
  \none & \none[] 
\end{ytableau},
 \begin{ytableau}
    \none[2]  &   & \\
 \none & \none[] & \none[] 
\end{ytableau}    \quad  \Big) \rightarrow 
\begin{ytableau}
    \none[1]  &   \\
    \none[1]  &   \\
  \none & \none[] 
\end{ytableau} 
$
\\
$
\Big(\begin{ytableau}
    \none[2]  &   & \\
 \none & \none[] & \none[] 
\end{ytableau}, 
\begin{ytableau}
    \none[1]  &   \\
    \none[1]  &   \\
  \none & \none[] 
\end{ytableau}  \quad 
    \Big) \rightarrow 
\begin{ytableau}
    \none[1]  &   \\
    \none[1]  &   \\
  \none & \none[] 
\end{ytableau} 
$
\\$
\Big(\begin{ytableau}
    \none[2]  &   & \\
 \none & \none[] & \none[] 
\end{ytableau},
\begin{ytableau}
    \none[2]  &   & \\
 \none & \none[] & \none[] 
\end{ytableau} \quad 
    \Big) \rightarrow  
\begin{ytableau}
    \none[2]  &   & \\
 \none & \none[] & \none[] 
\end{ytableau} 
$
\end{tabular}
\end{center}
\end{figure}

The  mechanism  for the example is in Table \ref{mappingR}.  The first column identifies a conflict in chosen   coalition structures. The second is a mapping from  strategies, chosen by players into strategies, after an application of the mechanism, in bold font. The last column is  resulting coalition structure, in bold font.  Equilibria strategies profiles are marked with a star, as usual.

 A result of the mechanism is a coalition structure:  an allocation of players over coalitions,  and strategies of players for this coalition structure.  A mechanism  $\mathcal{R}$  makes every emerging coalition structure be a playable non-cooperative game.  Table \ref{mappingR} is an explicit link  between Tables 2 and 3 to construct  coalition structures.

 \begin{center}
\begin{table}[p]
\caption{The rule:  the coalition structure $\{ 1,2 \}$ can be formed only from unanimous agreement of  both agents.The first column in an indicator of a conflict in choices of coalition structures, the second   is a mapping of initial strategies  of both players to the final strategies, the third is a final coalition. We use different fonts to discriminate strategies chosen from strategies formed by the mechanism. Equilibria  strategy profiles are marked with `*`. }
\label{mappingR}
\begin{tabular}{c|c|c}  \\
conflict & mapping  $\mathcal{R}(2) \colon S(K=2) \mapsto \textbf{S}(K=2)$ & \begin{tabular}{c}  final \\ coalition \\ structure \end{tabular} \\ \hline 
No &$({L}_{i,K=2, P_{separ} } , {L}_{-i,K=2, P_{separ} } )   \mapsto (\textbf{L}_{i,K=2, \textbf{P}_{separ} } , \textbf{L}_{-i,K=2, \textbf{P}_{separ} } )   $ & $\textbf{P}_{separ} $\\
No &$({L}_{i,K=2, P_{separ} } , {H}_{-i,K=2, P_{separ} } )   \mapsto  (\textbf{L}_{i,K=2, \textbf{P}_{separ} } , \textbf{H}_{-i,K=2, \textbf{P}_{separ} } ) $ & $\textbf{P}_{separ} $\\
No &$({H}_{i,K=2, P_{separ} } , {L}_{-i,K=2, P_{separ} } )   \mapsto (\textbf{H}_{i,K=2, \textbf{P}_{separ} } , \textbf{L}_{-i,K=2, \textbf{P}_{separ} } )$  & $\textbf{P}_{separ} $ \\
No & $({H}^*_{i,K=2, P_{separ} } , {H}^*_{-i,K=2, P_{joint} } )   \mapsto   (\textbf{H}^*_{i,K=2, \textbf{P}_{separ} } , \textbf{H}^*_{-i,K=2, \textbf{P}_{joint} } )  $ & $\textbf{P}_{separ} $ \\ \hline  \hline 
Yes &  $({L}_{i,K=2, P_{separ} } , {L}_{-i,K=2, P_{joint} } )   \mapsto  (\textbf{L}_{i,K=2, \textbf{P}_{separ} } , \textbf{L}_{-i,K=2, \textbf{P}_{separ} } )  $  & $\textbf{P}_{separ} $ \\
Yes &  $({H}_{i,K=2, P_{separ} } , {L}_{-i,K=2, P_{joint} } )   \mapsto  (\textbf{H}_{i,K=2, \textbf{P}_{separ} } , \textbf{L}_{-i,K=2, \textbf{P}_{separ} } )  $  & $\textbf{P}_{separ} $ \\
Yes & $({L}_{i,K=2, P_{separ} } ,  {H}_{-i,K=2, P_{joint} } )   \mapsto (\textbf{L}_{i,K=2, \textbf{P}_{separ} } , \textbf{H}_{-i,K=2, \textbf{P}_{separ} } )  $ & $\textbf{P}_{separ} $  \\
Yes & $({H}^*_{i,K=2, P_{separ} } , {H}^*_{-i,K=2, P_{joint} } )   \mapsto  (\textbf{H}^*_{i,K=2, \textbf{P}_{separ} } , \textbf{H}^*_{-i,K=2, \textbf{P}_{separ }} ) $  & $\textbf{P}_{separ} $ \\
\hline
Yes &$({L}_{i,K=2, P_{joint} } , {L}_{-i,K=2, P_{separ} } )   \mapsto  (\textbf{L}_{i,K=2, \textbf{P}_{separ} } , \textbf{L}_{-i,K=2, \textbf{P}_{separ} } ) $ & $\textbf{P}_{separ} $ \\
Yes &$({H}_{i,K=2, P_{joint} } , {L}_{-i,K=2, P_{separ} } )   \mapsto  (\textbf{H}_{i,K=2, \textbf{P}_{separ} } , \textbf{L}_{-i,K=2, P_{separ} } )$ & $\textbf{P}_{separ} $ \\
Yes &$({L}_{i,K=2, P_{joint} } , {H}_{-i,K=2, P_{separ} } )   \mapsto  (\textbf{L}_{i,K=2, \textbf{P}_{separ} } , \textbf{H}_{-i,K=2, P_{separ} } )  $ & $\textbf{P}_{separ} $ \\
Yes &$({H}^*_{i,K=2, P_{joint} } , {H}^*_{-i,K=2, P_{separ} } )   \mapsto (\textbf{H}^*_{i,K=2, \textbf{P}_{separ} } , \textbf{H}^*_{-i,K=2, P_{separ} } )    $ & $\textbf{P}_{separ} $ \\
\hline \hline
No & ${L}_{i,K=2, P_{joint} } , {L}_{-i,K=2, P_{joint} } )   \mapsto (\textbf{L}_{i,K=2, \textbf{P}_{joint} } , \textbf{L}_{-i,K=2, \textbf{P}_{joint} } )  $ & $\textbf{P}_{joint} $ \\
No & $({L}_{i,K=2, P_{joint} } ,{H}_{-i,K=2, P_{joint} } )   \mapsto  (\textbf{L}_{i,K=2, \textbf{P}_{joint} } , \textbf{H}_{-i,K=2, \textbf{P}_{joint} } ) $ & $\textbf{P}_{joint} $ \\
No & $({H}_{i,K=2, P_{joint} } ,{L}_{-i,K=2, P_{joint} } )   \mapsto (\textbf{H}_{i,K=2, \textbf{P}_{joint} } , \textbf{L}_{-i,K=2, \textbf{P}_{joint} } )  $  & $\textbf{P}_{joint} $ \\
No & $({H}^*_{i,K=2, P_{joint} } ,{H}^*_{-i,K=2, P_{joint} } )   \mapsto  (\textbf{H}^*_{i,K=2, \textbf{P}_{joint} } , \textbf{H}^*_{-i,K=2, \textbf{P}_{joint} } ) $  & $\textbf{P}_{joint} $ \\
\end{tabular}
\end{table}
\end{center}

There are  three ways to reach  equilibrium in $P_{separ}$, but only one for $P_{joint}$.
We can see, that in an equilibrium we need to assign probability 3/4 to $\textbf{H}^*_{i,K=2, P_{separ} }$, and  probability 1/4 to $\textbf{H}^*_{i,K=2, P_{joint} }$. So,  the coalition structure  $P_{separ}$  has equilibrium probability $3/4$, the coalition structure $P_{joint}$ has an equilibrium  probability $1/4$. However, probabilities for  the initial  strategies  are different, $Prob (S_{i,K=2,P_{separ}}) = Prob(S_{i,K=2,P_{joint}} = 2 ) = 1/2$, and we can  use them to construct  probabilities for the mechanism. The mechanism operates in a predetermined way, it loads strategies of the players, transforms them following some rule, and generates an outcome.

 Every  equilibrium is inefficient, and players have  negative  inter- or intra-coalition externalities, depending on which coalition structure is formed.   One need to note, that   for  $K=1$, there are no changes in construction of the game, and  a standard non-cooperative game of Nash is valid. But the difference is that here players have explicit coalitions of size one.
  
\section{Formal setup  of the  model}
 
Nash (1950, 1951) suggested a non-cooperative game that consists of a set of players $N$, with a general element $i$, a set of individual finite strategies $S_i,  i \in N$, and payoffs, defined as a mapping from a set of all strategies (possibly mixed) into a payoff profile of all players,  $S=\times_{i \in N} S_i \mapsto U=\Big(U_i (S)\Big)_{i \in N} \subset \mathbb{R}^{\#N}$, $ U_i < \infty,  i \in N$. An equilibrium in the game is a fixed point of the mapping   $S \mapsto S$,   with a relevant payoff profile $U^* = (U^*_1, \ldots, U^*_N)$.

For this paper we do not need   assumptions of 
CGT,  like  a number of players,  a number of deviators, threats from a  deviating coalition, etc to construct   an equilibrium.  We construct a model as a non-cooperative game.

 \subsection{Coalition structures and  strategies}
 Let  integer $N$, $2 \le N$, be a set of players and a number of players. Let a game   has  an integer parameter $K$, $1 \le K  \le N$, a maximum coalition size.  We will construct coalition structures with restrictions on a maximum coalition size  using Young diagrams. They describe collections of boxes from $N$ items with a restriction on the largest length $K$ of a  horizontal line of boxes (a ribbon in terms of the Young diagram literature). Exactly   one player  is  assigned to  one box,  a ribbon  is a coalition, and a length of a  ribbon is a coalition size, all ribbons describe coalition sizes, that partition $N$. 
Let $\mathcal{P(K)}$ be all possible Young diagrams from $N$ elements with the restriction $K$:  $$\mathcal{P}(K) = \Big\{P= \{g\} \colon g \subset N,  \forall \bar{g},  \tilde{g} \colon  \bar{g} \neq \tilde{g}, \cup g = N  \colon  \#g \le K \Big\}, $$ where $g$ is a coalition for a subset of players from $N$;    $\{g\}$ is a partition of $N$, or a coalition structure; different coalitions do not overlap, but together make the set of all players $N$. A size of any coalition does not exceed given $K$.    The same in terms of the Young diagrams: $g$ is a ribbon, length of a ribbon is no greater than $K$,  a sum of all ribbons is $N$; all ribbons in one diagram make a partition of $N$ into coalitions.  Examples of   Young diagrams   $\mathcal{P(K)}$  for three players  and different $K$s is in Figure  \ref{Ydiag}:

\begin{figure}
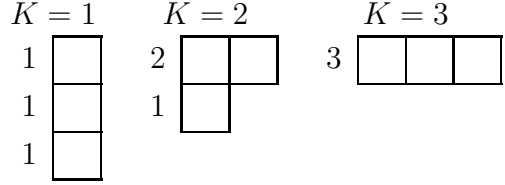

\caption{ Different  Young diagrams  with maximum coalition sizes $K=1,2,3$, and without allocation of players.}
\center
\label{Ydiag}
\begin{center}
 \begin{tabular}{ccc}
 $K=1$ &  $K=2$ & $K=3 $ \\
\begin{ytableau}
    \none[1]  &  \\
    \none[1]  &  \\
    \none[1]  &    \\
  \none & \none[] 
\end{ytableau}
&  \begin{ytableau}
    \none[2]  &   &  \\
    \none[1]  &   \\
  \none & \none[] 
\end{ytableau}
 &
 \begin{ytableau}
    \none[3]  &  & & \\
  \none & \none[] 
\end{ytableau}

 \end{tabular}.
\end{center}
\end{figure}

An increase in $K$ induces an inclusion of a Young diagram with a smaller $K$ by one with a greater $K$, what forms the Young lattice, a partially ordered set:  $$\mathcal{P}(K=1) \subset \ldots \subset \mathcal{P}(K) \subset  \ldots  \subset \mathcal{P}(K=N).$$  The bigger is $K$, the more  coalition structures are  involved into activity.  We have seen this in the example above.
  A power of  $\mathcal{P}(K)$ is a Bell number $B_K$ with a recurrence relation:
 $
 B_{K+1} = \sum^K_{k=0} \begin{pmatrix} K \\  k\end{pmatrix} B_k
 .$  Number of all possible allocations of players for all coalition structures is   $A^{B_K}_{N}$. 

\begin{figure}
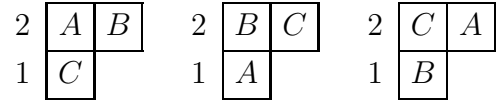

\caption{Different allocations of player   with the same structure of  Young diagrams. Players may have different strategy sets  in every case.}
\center
\label{Ydiags_allocs}
 \begin{center}
 \begin{tabular}{ccc}
 \begin{ytableau}
    \none[2]  &  A & B \\
    \none[1]  &  C  \\
  \none & \none[] 
\end{ytableau}
& 
 \begin{ytableau}
    \none[2]  &  B & C \\
    \none[1]  &  A  \\
  \none & \none[] 
\end{ytableau}
&
 \begin{ytableau}
    \none[2]  &  C & A \\
    \none[1]  &  B \\
  \none & \none[] 
\end{ytableau}
\end{tabular}
\end{center}
\end{figure}

  Allocations of players over boxes is the way to study externalities: a change in allocations is able to change  externalities. Thus we will discriminate Young diagrams  at Figure \ref{Ydiags_allocs}:  they  differ only by allocation of    players  over boxes, although they have similarities: one coalition size one, and one  coalition size two.  For  each of these cases a    player may have different strategy sets.
  
Let $P(K)$ be a diagram from $\mathcal{P}(K)$,   with fixed   sizes of coalitions. 
  Let  $\{ P(K)\}$ be all possible allocations of players  in  $P(K)$.  A player has a finite\footnote{Finite strategies are chosen as they used in  Nash (1950).} strategy set $S_i (P,K)$ for every coalition structure $P$ in $\{ P(K)\}$. Strategies may differ for different allocations of players, although a Young diagram is the same, see Figure \ref{Ydiags_allocs}.  A set of all strategies of $i$ is %
 \[ S_i(K) = \Big\{S_i(P, K) \colon  i \in N,  P \subset \{P(K)\}, P(K) \subset \mathcal{P}(K)\Big\},  \] or 
 $$ 
 S_i(K)=  \cup_{P(K) \in \mathcal{P} (K)}  \Big(\cup_{P \in \{P(K) \}} S_i (P,K) \Big)
,$$  where 
 $S_i(P, K)$ is a set of  finite strategies of $i$  with the maximum coalition  size  $K$ in the Young diagram $P$; $\{P (K)\}$ is a set of all allocations of players for a   coalition partition $P(K)$ with fixed  sizes of coalitions,   $P(K) \subset \mathcal{P}(K)$.    If  possible, we will write $S_i(P)$ for a particular $P$, omitting maximum coalition size $K$.
 
 Let  a  set  of all strategies of all players be $S(K)=\otimes_{i \in N} S_i (K)$, a tensor product of individual strategy sets of all players, where $K$ is a maximum coalition size. 
  Individual  strategy sets for different $K$ are    nested:
 $$S_i(K=1)  \subset \ldots \subset S_i(K) \subset  \ldots  \subset  S_i(K=N),$$  what we have observed in the example above.

 A strategy set of all players is constructed as it was in the example of the previous section:
 $$
 S(K) =  \cup_{P(K)  \in \mathcal{P}(K)\}} \Big(\cup_{P \in \{P(K) \}}  \Big(\otimes_{i \in N}  S_i (P,K) \Big) \Big)
 .$$
 So far we have expanded the strategies setup of a standard non-cooperative game  to a case, with coalition structures.  
 ction.

\subsection{Mechanism}


Coalition structures  chosen by  different   players can be incompatible, we have seen this above. In practice, such  conflicts are resolved by  social norms,  customs and traditions,  brute force and enforcements along with a variety of   different transaction costs and imperfect information. These social artifacts we will address as a social mechanism.  Individual impact  of a mechanism is  that  an individual  choice of a coalition structure  may not come true.

A mechanism determines  final alternatives for  each player, where one can deviate. This makes an individual choice be bounded, bounded by the environment. The alternatives are enumerated by final coalition structures and allocations of players over them.

A social mechanism joins  a strategy profile from   $N$ players and transforms it into  one final coalition structure, and a  relevant strategy  profile for this coalition structure.   
Let  $\mathcal{R}(K)$ be a mechanism, which transforms  every    strategy profile  $(s_1,\ldots, s_i,\ldots,s_N)$, $s_i \in S_i(K)$,  $i \in N$,   into    some  strategy set of a mechanism: $$\textbf{S}( K) = \cup_{ \textbf{P}(K)}  \Big(\cup_{\textbf{P} \in \{\textbf{P} (K) \}}  \otimes_{i \in N}S_i(\textbf{P}) \Big),$$ where a final coalition structure $\textbf{P}$ serves as an index.\footnote{This means, a Young diagram fixes sizes of coalitions,  and we need to list allocations of  all players over  relevant coalitions.}  Let ${\mathbf{P}}(K)$ be all coalition structures, supported by $\mathcal{R}(K)$, and let $\{\textbf{P} (K) \}$ be the implementable by the mechanism coalition structures, which differ by allocation of players, and $S_i(P)$ is a set of strategies of $i$ for a coalition structure $P$. In the example below, from three  diagrams in  Figure  \ref{Ydiags_allocs}, a mechanism     supports formation of only one.

We   assume, that a strategy  profile in $S(K)$ is transformed in a strategy profile of $\textbf{S}(K)$, but not vice verse.  
 We continue the example of three players, let $K=2$, and  final coalition structures are: \begin{ytableau}
    \none[1]  &  A    \\
    \none[1]  &  B   \\
    \none[1]  &   C \\
  \none & \none[] 
\end{ytableau}  and \begin{ytableau}
    \none[2]  &  C &  A  \\
    \none[1]  &  B  \\
  \none & \none[] 
\end{ytableau}, see   Table \ref{mechExample}, where    $\textbf{S}_i (\textbf{P},K=2)$  is a strategy set of $i$ after application of the mechanism. Note, that  even $\begin{ytableau}
    \none[2]  &  C &  A  \\
    \none[1]  &  B  \\
  \none & \none[] 
\end{ytableau}$   is formed, there is  $$S_i \Big(\begin{ytableau}
    \none[2]  &  C &  A  \\
    \none[1]  &  B  \\
  \none & \none[] 
\end{ytableau}, K=2 \Big) \neq \textbf{S}_i \Big(\begin{ytableau}
    \none[2]  &  C &  A  \\
    \none[1]  &  B  \\
  \none & \none[] 
\end{ytableau} , K=2\Big).$$  
  For this example we do not  need details, how to construct $\textbf{S}_i(P)$.

\begin{table}[p]
\caption{Example of a mechanism  to form two coalition structures  from three agents $A,B,C$ and $K=2$. Strategy set of $i$ in a final coalition structure $\textbf{P}$ is $\textbf{S}_i (\textbf{P},K=2)$. The mechanism of the example supports formation of only two coalition structures.  Note, that there are three coalition structures of the same type
, but we do not  rank  them;  this will complicate the exposition, but will not change the idea. There are mixed strategies, which will be used for  expected utility construction further. Every time only one coalition structure can be realized.}

\begin{center}
\begin{tabular}{c||c|c|c|c}
\begin{tabular}{c} Initial \\ Young diagrams\\
$\mathcal{P} (K=2)$
\end{tabular}  & \begin{ytableau}
    \none[1]  &     \\
    \none[1]  &     \\
    \none[1]  &    \\
  \none & \none[] 
\end{ytableau}
 & \begin{ytableau}
    \none[2]  &   &   \\
    \none[1]  &    \\
  \none & \none[] 
\end{ytableau} 
& & 
 \\ \hline
\begin{tabular}{c}  Possible \\ allocations \\ of players \\
$P \in  $ \\ $\{P (K=2)\}$
\end{tabular}  & \begin{ytableau}
    \none[1]  &  A    \\
    \none[1]  &  B   \\
    \none[1]  &   C \\
  \none & \none[] 
\end{ytableau} 
&\begin{ytableau}
    \none[2]  & A  &  B \\
    \none[1]   & C    \\
  \none & \none[] 
\end{ytableau} & 
\begin{ytableau}
    \none[2]  &  C &  A  \\
    \none[1]  &  B  \\
  \none & \none[] 
\end{ytableau}
&
\begin{ytableau}
    \none[2]  &  B &  C  \\
    \none[1]  &  A  \\
  \none & \none[] 
\end{ytableau}

 \\ \hline
\begin{tabular}{c}    Coalition \\ structures \\ implementable  \\ by  
$\mathcal{R}(K=2)$, \\
$\textbf{P} \in \{\textbf{P} (K) \}$. 
\end{tabular}  & 
\begin{tabular}{c}
\textbf{P}=\begin{ytableau}
    \none[1]  &  A    \\
    \none[1]  &  B   \\
    \none[1]  &   C \\
  \none & \none[] 
\end{ytableau}  
\end{tabular}
& - &  \begin{tabular}{c}
\textbf{P}=\begin{ytableau}
    \none[2]  &  C &  A  \\
    \none[1]  &  B  \\
  \none & \none[] 
\end{ytableau}  
\end{tabular}
& -

\\ \hline
\begin{tabular}{c}
Coalition structure \\
 strategies  set  \\ 
 $\textbf{S} (K=2)  $ \\ $= \cup_{\textbf{P}} \textbf{S}_{\textbf{P}}$
 \end{tabular} & \begin{tabular}{c} $\textbf{S}_{\textbf{ P}} =  $  \\ $\times_{i \in \atop{\{A,B,C\}}} $   \\ $ \textbf{S}_i (\textbf{P},K=2)$ \end{tabular}   &  -
& \begin{tabular}{c} $\textbf{S}_{\textbf{ P}} =  $  \\ $\times_{i \in \atop{\{A,B,C\}}} $   \\ $ \textbf{S}_i (\textbf{P},K=2)$ \end{tabular}  & -
\\ \hline
\begin{tabular}{c}
Individual mixed \\ strategies  
for a   \\
coalition structure, \\
$\sum_{\textbf{P}} \Sigma_i (\textbf{P},K) =1$
\end{tabular} &
$ \Big(\Sigma_i (\textbf{P})\Big)_{i \in {A,B,C}}$ &  -  & $  \Big(\Sigma_i(\textbf{P})\Big)_{i \in {A,B,C}}$ & - 
 \\ \hline
\begin{tabular}{c}
Individual  \\ payoffs
\end{tabular}
 &
 $\Big(U_i (\textbf{P}, \textbf{S}_i) \Big)_{i  \in \{A,B,C\}}$
 &
-
 &
 $\Big(U_i (\textbf{P}, \textbf{S}_i) \Big)_{i  \in \{A,B,C\}}$  & - \\  \hline
 
\end{tabular}
\end{center}
\label{mechExample}
\end{table}%

\begin{definition}  \textbf{A coalition structure formation mechanism}  $\mathcal{R}(K)$  is a mapping such that:
 \begin{enumerate}
\item Domain of $\mathcal{R}(K)$  is a set of all strategy profiles,   $S(K)$. 
\item A range of $\mathcal{R}(K)$   is
\begin{itemize}
\item a set of final (implementable) coalition partitions  ${\mathbf{P}}(K) \subset \mathcal{P}(K),$ with a general element 
$\textbf{P}$, 
\item 
  a set of strategies  $\textbf{S} \Big(\textbf{P}(K),K\Big)$ with a general element 
 $\textbf{s} = (\textbf{s}_1, \ldots, \textbf{s}_N)$, $\textbf{S}(K) \subset S(K)$.
\end{itemize}
 
\end{enumerate}
 \end{definition}

The mechanism $\mathcal{R}(K)$ reframes $S(K)$ into   implementable, and non-implementable  coalition structures, along with  relevant  sets of strategies for implementable: 
$ \textbf{S}(K) =  \cup_{\textbf{P}(K)} \Big(\cup_{ \textbf{P} \in \{ \mathbf{P}(K)\} } \textbf{\textbf{S}}\Big(\textbf{P},K\Big) \Big)$, 
where $\textbf{S}(\textbf{P},K) = \times_{i \in N} S_i (\textbf{P}, K)$ is  set of strategies  of all players for a coalition structure $P$.  
   Two different  final coalition structures, $\bar{\mathbf{P}}$ and $\tilde{\textbf{P}}$, $ \bar{\textbf{P}} \neq \tilde{\textbf{P}}$,  have  different coalition structure   strategy sets    $  \textbf{S}( \bar{\textbf{P}},K) \cap \textbf{S}(\tilde{\textbf{P}},K) = \emptyset$.
   
Formally the same:
 \begin{multline*}\mathcal{R}(K) \colon 
S(K)=\otimes_{i \in N}S_i(K)  \mapsto   \colon
 \begin{cases} 
 \textbf{S}(K) \subseteq S(K), \\
 \textbf{S}(K) =   \cup_{\textbf{P}(K)} \Big(\cup_{\textbf{P} \in \{\mathbf{P} (K) \} }\textbf{\textbf{S}}(\textbf{P},K) \Big), \\
  \forall \bar{\textbf{P}},\tilde{\mathbf{P}} \in  \{ \mathbf{P}(K)\} \subseteq \textbf{P}(K),  \bar{\textbf{P}} \neq \tilde{\textbf{P}} \Rightarrow \textbf{S}( \bar{P},K) \cap \textbf{S}(\tilde{P}, K) = \{\emptyset \} \\
 \forall  s = (s_1,...,s_N) \in S(K)  \\  \mbox{        } \exists \textbf{s}=(\textbf{s}_1,\ldots,\textbf{s}_N)  \in \textbf{S}(K)  \\  \mbox{     such that }  \,   \exists \textbf{P}  \colon
 \textbf{s} \in \textbf{S}(\textbf{P},K),    \\
 \end{cases} .
\end{multline*} 

We assume, that even if  resulting  $\textbf{S}( \bar{\textbf{P}},K)$  and   $ \textbf{S}(\tilde{\textbf{P}}, K)$ are numerically identical, but  correspond to different coalition structures, $\bar{\textbf{P}} \neq \tilde{\textbf{P}}$, we  consider them as  different,   $\textbf{S}( \bar{\textbf{P}},K)  \neq \textbf{S}(\tilde{\textbf{P}}, K)$. 
We also assume that  $\mathcal{R}(K) $ is a predetermined mechanism, it does not randomize, and it is known for everybody. 

\begin{definition}
 The family of mechanisms with 
 $$\mathscr{R} = \mathcal{R}(K=1)  \subset \ldots \subset \mathcal{R}(K)  \subset  \ldots \subset \mathcal{R}(K=N)
 ,$$ is a nested family of rules.
\end{definition}

This means just consistency in formation of coalition structures, if   maximum coalition size increases.  It is not a necessary restriction, but in many cases it  may be desirable.

One may think about the game as a two-stage game: first   players  move, then a mechanism. But as far a mechanism is known for the players, this is a non-necessary complication.  


\subsection{Payoffs}
The next step is to construct coalition partition specific payoffs.  
 For every  final coalition  structure  $\textbf{P}$  player $i$ has a payoff function $$U_i(\textbf{P}, K) \colon  \textbf{S}(\textbf{P},K) \rightarrow \mathbb{R}_+,$$ such that   every  $U_i(\textbf{P},K)$ is  bounded, $U_i(\textbf{P},K)  < \infty$,  and $\textbf{S}(\textbf{P},K)$ is a set of feasible strategies of all players for $\textbf{P}$.  We construct payoffs only on coalition structures implementable by a mechanism.

   Every coalition structure   has an  allocation of  players, every player has a set of strategies for this coalition structure and a payoff. Thus every coalition structure is a non-cooperative game, and it is a playable game.  Each time only one such game is played . If a game, or a realization of a coalition structure, is played,   all  players are exposed for intra and inter-coalition externalities, which depend on allocation of players  and strategies of the players in this coalition structure.

Let  $$\mathcal{U}_i(K)  = \Big\{U_i(\textbf{P}, K) \colon \textbf{S}(\textbf{P},K) \subset \textbf{S}(K), \textbf{P} \in \{ \mathbf{P}(K)\}  \subset \mathbf{P}(K) \subseteq \mathcal{P}(K) \quad | \quad  \mathcal{R}(K) \Big\} $$ be a set of all  payoffs of $i$ for all implementable coalition 
structures  with all possible allocations of  the players. 

 An increase in $K$ increases the number of possible partitions and the set of strategies. Hence   we obtain a nested family of individual payoff functions:
 $$ \mathcal{U}_i(K=1) \subset  \ldots \subset  \mathcal{U}_i(K) \subset  \ldots \subset \mathcal{U}_i(K=N).$$ 

  We have observed the nested property of payoffs in the matrix game above. 
  \subsubsection{A game}
  
\begin{definition}[\textbf{a simultaneous  non-cooperative coalition structure formation game}]
A non-cooperative game   for coalition structure formation consists of two parts, an environment and  properties of agents: 
$$\Gamma(K)=\Big\langle  \underbrace{\Big\{K, \mathcal{P}(K),\mathcal{R}(K) \Big\}}_{\mbox{environment}}, \underbrace{\Big(N, S_i(K),\mathcal{U}_i (K)\Big)_{i \in N}}_{\mbox{properties of agents}} \Big\rangle, $$ where 
  
\begin{itemize}
\item $K$  is a maximumm coalition size,
\item $\mathcal{P}(K)$ is a set of Young diagrams for $N$ players,
\item $\mathcal{R}(K)$ is a coalition structure formation mechanism,
\item $N$  is a set of agents with a general element $i$, $N \ge K$,
 \item ${S}_i(K)$ is a set of individual strategies of $i$,
\item $\mathcal{U}_i (\textbf{P}, K)$ - is a set of payoffs, defined over coalition partitions, supplied by the mechanism $\mathcal{R}$,

\end{itemize}

such that: 
 $$ \Big( \mathcal{P}(K), S(K)\Big) \buildrel \mathcal{R}(K) \over \rightarrow \Big\{ \Big(\textbf{P}, \textbf{S}(\textbf{P},K) \Big) \colon \textbf{P} \in \{\textbf{P(K)} \} \subset \mathbf{P}(K) \Big\} \rightarrow \Big\{ \Big(\mathcal{ \textbf{P}, U}_i(\textbf{P}, K)\Big)_{i \in N \atop{\textbf{P} \in \{\textbf{P(K)} \} \subset \mathbf{P}(K) }}\Big\}. $$ 
\end{definition}
For $K=1$  we obtain the traditional non-cooperative game of Nash:  $S=\times_{i \in N} S_i  \rightarrow  \Big(U_i(S,K=1)\Big)_{i \in N} $, but with  explicit allocations of players in coalitions size 1.

Varying $K$,  a maximum coalition size, we obtain a family of the nested games with the trivially proved proposition.

\begin{proposition}[\textbf{family of games}]
A family of the games:
$$
\mathcal{G}=\Gamma(K=1) \subset \ldots \subset \Gamma(K) \subset \ldots \subset \Gamma(K=N).
$$
is \textbf{ nested} for a nested family of mechanisms,
$$\mathscr{R} = \mathcal{R}(K=1)  \subset \ldots \subset \mathcal{R}(K)  \subset  \ldots \subset \mathcal{R}(K=N)
 .$$
 \end{proposition}

A game of finite strategies may have no pure strategies equilibrium. 
 Further   we assume that the mechanism $\mathcal{R}(K) $ does not randomize. 

 \subsection{Mixed strategies}
 We can define mixed strategies over initial strategy set  $(S_1, \ldots, S_N)$ of all players,   or  a strategy  set $(\textbf{S}_1, \ldots, \textbf{S}_N)$ implementable by the mechanism. We follow the second way, see Table \ref{mechExample}. 
 Let $\Sigma_i (K)$ be a set of all mixed strategies  of $i$,  defined over $\textbf{S}_i(K)$: $$\Sigma_i (K) =\Big\{\sigma_i (K) \colon \int_{\textbf{S}_i(\textbf{P}, K) \atop {\forall \textbf{P}}} d\sigma_i(K)= 1 \Big\},$$  with a general element $\sigma_i (K)$, where,   $\textbf{S}_i (\textbf{P} ,K)$ is a strategy set  of $i$ implementable   by $\mathcal{R}(K)$ for a coalition structure $\textbf{P}$.  Normalizing integral is the Lebegue integral defined over all implementable coalition structures $\textbf{P}$. The set  $\Sigma_i (K)$ is convex by properties of probability spaces. For simplicity,  we do not use bold font for mixed strategies, although they are defined over strategy sets after application of the mechanism $\mathcal{R}(K)$. 
 
 We have observed  mixed strategies in the example  in Table \ref{default2}, where in  equilibrium  the mechanism  randomizes  between $\textbf{H}_i(K=2,P_{separ})$ and $\textbf{H}_i(K=2,P_{joint})$,  or players  equally randomizing between   ${H}_i(K=2,P_{separ})$ and ${H}_i(K=2,P_{joint})$.   
  
  
Expected utility of $i$  is defined  as usual:
$$
EU^{\Gamma(K)}_i \Big(\sigma_i(K),\sigma_{-i}(K)\Big) = \int_{
  \textbf{S}(K) = \cup \textbf{S}(\textbf{P},K)} 
   {U}_i(\textbf{s}_i,\textbf{s}_{-i}) d\sigma_i(K) d\sigma_{-i}(K).$$
 
Expected utility is constructed in the standard way: a weighted payoff for every possible outcome of the game, while an outcome includes a coalition structure.
 
  At the moment we have all components to define an equilibrium in mixed strategies with formation of coalition partitions. 
  
 \begin{definition}[\textbf{an equilibrium}]
A mixed strategies profile  $\sigma^*(K)=\Big(\sigma^*_i(K)\Big)_{i \in N}$ is  an equilibrium strategy profile for a game $\Gamma(K)$ if for every  
  $\sigma_i(K) \neq \sigma^*_i(K)$ 
the following inequality for every player $i$ from $N$ holds true:
 $$
 EU^{\Gamma(K)}_i \Big(\sigma^*_i(K),\sigma^*_{-i}(K)\Big) \ge EU^{\Gamma(K)}_i \Big(\sigma_i(K),\sigma^*_{-i}(K)\Big).
  $$
\end{definition}

\begin{proposition}
An equilibrium in mixed strategies always exists.
\end{proposition}

A proof follows from topological considerations of convexity of probability spaces defined over finite tensors spaces.

Equilibrium in the game  $\Gamma (K)$ is defined in the standard way,  it is just an expansion of Nash theorem.
The existence result is different from  those of  CGT approach, where an equilibrium may not exist. There is no bargaining, and there is complete information about fundamentals of the  game. Another outcome of the model  is  that there is no need to introduce additional properties of games, like axioms on a system of payoffs, super-additivity, weights, etc.  The  goal of cooperative game theory is achieved within a non-cooperative framework. 
Equilibrium existence result  can be generalized for the whole family of games.
  
\begin{theorem}
The family of games $\mathcal{G}=\{ \Gamma(K), K=1,2, \dots,N\}$ has  an equilibrium  in mixed strategies,  $$\sigma^*(\mathcal{G})=\Big(\sigma^*(K=1), \ldots,\sigma^*(K=N) \Big)$$
\end{theorem}
This result is obvious,  and it expands the classical Nash theorem.  Actually it says  nothing about inheritance  of properties in the  set of mechanisms $\mathcal{R}(K)$, $K = 1,N$. 
An equilibrium in the game can also be  characterized  by   equilibrium partitions. 
\begin{definition}[\textbf{equilibrium coalition structures or partitions}]
A set of  partitions $\{ \textbf{P}^*\}(K)$, $\{ \textbf{P}^*\}(K) \subset \mathcal{\textbf{P}}(K)$, of a game $\Gamma(K)$, is a set of equilibrium (mixed) partitions,  if it is induced by an equilibrium strategy profile $\sigma^*(K)=\Big(\sigma^*_i(K)\Big)_{ i \in N}$. 
\end{definition}

We need to note, that even if a coalition structure is implementable by $\mathcal{R}(K)$, that  does not imply, that  every implementable coalition structure is  in an equilibrium.  We also assume that  a mechanism is free, and there are  no costs to form coalition structures. These complications  require introduction of resources into the game.

 \section{Stability}A desirability of an individual deviation depends on individual expectation from every feasible alternative to deviate. Here,  a player has alternatives, described in terms of  formed coalition structures and  allocations of other players over them.  And the Young diagrams serve to enumerate  the appearing alternatives.   This resolves the problem of recurrent  formation of coalitions of deviators themselves  implicitly hidden in  approaches based on threats and inside/outside deviators. 
 
There are many ways to think about  alternatives and induced stability within the framework of the model. We  limit the scope  only to stability in terms of a  maximum coalition size $K$, what  means that we are working with a framework of the whole family of games $\mathcal{G}$. Every criterion induces corresponding stable partition structures, which are  omitted   here. Stability is understood  in terms of mixed strategies.  We introduce local, global   stability  criteria, and a criterion for the strong Nash equilibrium.

 \begin{definition}[local non-cooperative coalition structure stability criterion]
 The family of games $\mathcal{G}=\{ \Gamma(K), K=1,2, \dots,N\}$ has
a mixed strategies profile  $\sigma^*(K)=\Big(\sigma^*_i(K)\Big)_{i \in N}$  as  a   locally stable  profile if for every   game $\Gamma(K)  \in \mathcal{G}$ and 
  $\sigma_i(K) \neq \sigma^*_i(K)$, $i \in N$,  there is 
$$
 EU^{\Gamma(K)}_i \Big(\sigma^*_i(K),\sigma^*_{-i}(K)\Big) \ge EU^{\Gamma(K_1)}_i \Big(\sigma_i(K_1),\sigma^*_{-i}(K_1)\Big),
  $$
  and  $K_1  \in \{\max{0, K-1} , \min \{K+1, N\}\}$.
\end{definition}
This means  that one unit deviation in coalition size  can  not deteriorate expected  payoffs.   
 A locally stable mixed strategy equilibrium profile induces corresponding  locally stable coalition structures.  
 
 The global  criterion    is constructed in the natural way: there is only one $K$, when a family of games induces maximum expected payoffs for  each player. 
  
  \begin{definition}[global  non-cooperative coalition structure stability criterion]
 The family of games $\mathcal{G}=\{ \Gamma(K), K=1,2, \dots,N\}$  has
a mixed strategies profile  $\sigma^*(K)=\Big(\sigma^*_i(K)\Big)_{i \in N}$ as a globally stable  profile  if  for every   game $\Gamma(K)  \in \mathcal{G}$, for every  
  $\sigma_i(K) \neq \sigma^*_i(K)$, 
   and  every  maximum coalition size $K_1$ different  from $K$, $ K_1 \neq K$ there is:
 $$
 EU^{\Gamma(K)}_i \Big(\sigma^*_i(K),\sigma^*_{-i}(K)\Big) \ge EU^{\Gamma(K_1)}_i \Big(\sigma_i(K_1),\sigma^*_{-i}(K_1)\Big),$$ 
  $i \in  N$.  
\end{definition} 
  
  And finally we can reconsider the strong Nash equilibrium in terms of stability of  the suggested non-cooperative game. The strong Nash equilibrium has a restriction on  the coalition size, it must  have the  maximum size. This means the same, as ``all agents in one coalition,''  but allows to make  more.

  \begin{definition}[non-cooperative strong Nash equilibrium criterion]
 The family of games $\mathcal{G}=\{ \Gamma(K), K=1,2, \dots,N\}$  has
a mixed strategies profile     $\sigma^*(K)=\Big(\sigma^*_i(K)\Big)_{i \in N}$ as a  strong mixed strategy profile if for every   game $\Gamma(K)  \in \mathcal{G}$,  for every  
  $\sigma_i(K) \neq \sigma^*_i(K)$,  $i \in N$, 
there is:
 \begin{itemize}
\item 
 $
 EU^{\Gamma(K)}_i \Big(\sigma^*_i(K),\sigma^*_{-i}(K)\Big) \ge EU^{\Gamma(K_1)}_i \Big(\sigma_i(K_1),\sigma^*_{-i}(K_1)\Big),$
 \item $ \forall K_1 \neq  K, K = N,$
 \item  and only the grand coalition is in an equilibrium, $ \{\textbf{P}(K=N) \} = \{ N\}$.

\end{itemize}

\end{definition}
This  criterion supports a non-cooperative  criterion  for the strong Nash equilibrium.  Aumann (1990) claimed in the title that ``Nash equilibria are not self-enforcing."  We provide a non-cooperative criterion, when    the grand coalition is self-enforcing.

 One can take this criterion as  ``the noncooperative implementation of the ... cooperative solution'' (Serrano, 2021), in the meaning of a cooperation for everybody.   The criterion induces the same result, as the strong Nash equilibrium, but it  explicitly deals with individual actions, not with inside or outside deviators. 

\section{Discussion}

 The justification of  a  non-cooperative game approach applied to  studying coalition formation, comes from   Maskin (2011) and a     remark of 
 Serrano (2014), 
 that for studying coalition formation \textquotedblleft it  may be worth  to use strategic-form   games, as proposed in the Nash program.\textquotedblright
 
 The recent survey of advances in solutions for the Nash Program is Serrano (2021),   but it   has a limited attention to non-cooperative toolkit for the problem.
 


Insufficiency of cooperative game theory to study  coalitions and coalition structures was earlier reported by many authors. Maskin (2011) wrote that \textquotedblleft  features of cooperative theory are problematic because most applications of game theory to economics involve settings in which externalities are important, Pareto inefficiency arises, and the grand coalition does not form.\textquotedblright  
  Myerson (p.370, 1991) noted that  \textquotedblleft  we need some model of cooperative behavior that does not abandon the individual decision-theoretic foundations of game theory.\textquotedblright Thus there is a  demand for   a specially designed  non-cooperative game to study coalition structures formation along  with an adequate equilibrium concept for this game.\footnote{Serrano (2021), describes the  challenge of the Nash Program in the following way: ``The initial interpretation of the Nash program, as formulated in Nash (1953), was to
describe the strategic rules of negotiation underlying an axiomatic solution. According
to this view, the primitive is a given axiomatic solution and the goal is to enhance its
understanding, by obtaining it as a result of a completely different approach.''  The model demonstrates, that coalitions can be formed as a non-cooperative one stage game with complete information. One can extend the game in time in the usual way.}
 The  research  program in Table ~1 satisfies these requirements.
 
 There is a voluminous literature on the topic, the short list of authors is far from complete:  Aumann, Hart, Holt,  Maschler,  Maskin, Myerson, Peleg,  Roth, Serrano, Shapley, Schmeidler, Trockel, Weber, Winter, Wooders and many others. Serrano (2021) presents statistics of papers on the topic.

Much work has been done, taking  a   \textquotedblleft threat\textquotedblright   as a basic concept for  coalition   formation analysis, as it was initially suggested by Nash (1953).    
   Consider a  strategy profile from  a subset of   players. Let this profile be  a threat to someone, beyond this subset.  
Threatening  and deviating player\textbf{s}  may  produce externalities   for each other (and negative  externalities not excluding!). Literature studies credibility of the threat based on the dilemma ``to join or not.''  In this paper players have much wider menu of actions.

Also, the deviators   may produce gains for someone in other coalitions, or they can join other coalitions, and produce other externalities for the coalition of an origin, and induce a chain of  further deviations.  And  finally the problem becomes recurrent and intricate.  

Blocking approach is another form of a threat, for example Aumann, Dreze, (1974), who studied cooperative approach to coalition structures.  The weak point here is the same as with a threat. Those who block,    can form many blocking coalitions, which can even neutralize mutual threat to the original coalition, and the story of a non-cooperative  coalition structure construction repeats. Actually,  further investigation of these kind of reasonings has led to  this paper.

Demange (2017)  provides a survey on group formation from coalition deviation. 
All decisions in the current model are made individually, before a  coalition structure is formed. Mobility between groups, individually based and rational,   may happen in the presented model  in  mixed strategies cases. Possibilities for   new groups to emerge depend on strategies and   intra and inter coalition externalities.

Serrano (2021) provides  a survey for ``bargaining design,'' what intrinsically deals with imperfect information and games in extended form. Our model can be  developed as a game in extended form too, with  multiple coalitions  in an equilibrium at every stage. We expect that all the  existing machinery of non-cooperative games can be applied for the presented toolkit.

The difference of the research agenda   of this paper  with  the  Nash program (Serrano 2004, 2021) is   the attention to   non-cooperative formation  of   complex coalition structures,   different from one grand coalition. The best analogy for the difference   is the  difference between partial  equilibrium at one market and general equilibrium  for many markets. The former isolates a market  ignoring cross-market interactions, the latter explicitly studies cross market interactions. This analogy is similar to intra and inter coalition externalities. 

The constructed model satisfies the  goals setup in Table \ref{differences}. 
The  finite non-cooperative game allows to  study  what can be a cooperative behavior, when  the individuals
 \textquotedblleft    rationally further their individual interests" (Olson, 2009).  

This paper has two contributions in comparison to the original paper  of Nash:  a  construction of a family of non-cooperative games with an  embedding  coalition structure formation mechanism, and stability criteria. The  non-cooperative game suggested by Nash is a partial case for these games.

  Every game in a family   has an equilibrium, may be in mixed strategies. This differs from results of cooperative game theory, where  games may have no equilibrium, like  games with empty cores, etc. 
 
 The introduced  equilibrium concept differs from the strong  Nash, coalition-proof (Bernheim, Peleg,   Whinston, 1987) and $k$-equilibrium concepts. The differences  are:   non-cooperative approach to coalition formation, an explicit allocation of payoffs and a combined presence of intra- and inter- coalition (or group) externalities. The list  of differences is not complete and can be specified for any relevant paper.  
 
 Differences from the core approach of Aumann (1960)  and from voluminous cooperative games literature are obvious:  a presence of externalities, no restrictions that  only one group deviates, no restrictions on the direction of a deviation (inside or outside),   a  construction of individual payoffs from a strategy profile of all players, an equilibrium always exists. 
The approach allows to  study    coalition structures, which differ  from the grand coalition  as in   Shapley value.  
 Finally the introduced concepts enables to offer  a set of non-cooperative stability criteria,  what is  not done in other literature.

The suggested  approach is different  in a role for a central planner offered by Nash, who \textquotedblleft  argued that cooperative actions are the result of some process of bargaining" Myerson (p.370, 1991).  
There are two possible understanding  for ``cooperation'' in this paper. First, allocation in one coalition. second. creating positive externalities, even being in different coalitions. 

Guido, Robbett and Romaniuc (2019) survey   cooperation for group formation  for social dilemma games. They claim that   a cooperation  more likely to emerge  endogenously for ``like-minded partners,"  than in ``exogenously formed groups."  The presented model can be used to respond for this kind of empirical research.  
 
\section{Conclusion}
 The paper suggests a new family of  simultaneous, non-cooperative, finite  games with the properties:   in an equilibrium players  can  stay  in different coalitions and be exposed for intra- and inter- coalition externalities.  The  model separates,  agents, and a social mechanism  to construct coalition structures.   The setup  of the research program in  Table 1 is fulfilled: the constructed game satisfies the required features,
 Using  the model  one can study  coalition structure stability in terms of non-cooperative games, with  the developed machinery  of non-cooperative games. 
The approach can be applied and further developed for   many areas of social sciences.  

\end{document}